# Simpler embeddings of causal sets into Minkowski spacetime

Steven Johnston [*]

We present a new method for embedding a causal set into Minkowski spacetime. The method is similar to a previously presented method, but is simpler and provides better embedding results. The method uses spacetime volumes to define causal set analogs of time coordinates for all elements, and spatial distances for pairs of causally related elements. The spatial distances for causally related pairs are then used to derive spatial distances for spacelike separated pairs by applying the triangle inequality. The result is a matrix of spatial distances for all pairs of elements in the causal set. This distance matrix can be decomposed to give coordinates in Minkowski spacetime. Results are presented showing good quality embeddings into Minkowski spacetime for dimensions $d = 2, 3, 4$.

## I. INTRODUCTION

Causal set theory provides a model in which spacetime is fundamentally discrete. Spacetime events are represented by elements of a causal set—a locally finite, partially ordered set in which the partial order represents the causal relationships between events in spacetime. The reader is directed to [1] for introductions, motivations and further references.

There has been extensive work on how to recover properties of continuum spacetimes starting with just a causal set. Recovering distances is straightforward for causally related pairs of elements, but more difficult for spacelike separated pairs. Past work looking at spacelike distances includes [2–4]. In addition to distances, there is also interest in finding embeddings of causal sets into continuum (often Minkowski) spacetimes, as described in [5–8].

We have previously presented a method [5] to embed a causal set into $d$-dimensional Minkowski spacetime $\mathbb{M}^d$. The current work presents a new simpler method that also gives better embedding results. To reduce duplication, we refer the reader to [5] for background and motivations.

## II. CAUSAL SETS AND MINKOWSKI SPACETIME

A "causal set" is a locally finite partially ordered set $(\mathcal{C}, \preccurlyeq)$ with a set $\mathcal{C}$ and a partial order relation $\preccurlyeq$ defined on $\mathcal{C}$. We shall label elements of $\mathcal{C}$ as $v_x$ for $x = 1, \ldots, |\mathcal{C}|$. We write $v_x \prec v_y$ to mean $v_x \preccurlyeq v_y$ and $v_x \neq v_y$. The set $\mathcal{C}$ represents the set of spacetime events and the partial order $\preccurlyeq$ represents the causal order between pairs of events. The spacetime volume $V$ of a region in a causal set is proportional to the number of elements $n$ in the region: $V = n/\rho$ for a volume density $\rho$.

If $v_x \preccurlyeq v_y$ we say "$v_x$ precedes $v_y$" and define their causal interval as all elements causally between them: $[v_x, v_y] \coloneqq \{v_z \in \mathcal{C} | v_x \preccurlyeq v_z \preccurlyeq v_y\}$. If two elements are unrelated (neither $v_x \preccurlyeq v_y$ or $v_y \preccurlyeq v_x$) we say they are spacelike separated and write $v_x || v_y$.

Points in $d$-dimensional Minkowski spacetime $\mathbb{M}^d$ are $d$-component vectors $x = (x_0, \mathbf{x})$ with $x_0 \in \mathbb{R}$ the time component and $\mathbf{x} = (x_1, \ldots, x_{d-1}) \in \mathbb{R}^{d-1}$ the space component. We write the Euclidean spatial norm as $|\mathbf{x}|$ and the squared Minkowski norm as

$$x^2 = x_0^2 - |\mathbf{x}|^2 = x_0^2 - \sum_{i=1}^{d-1} x_i^2. \quad (1)$$

A vector $x$ is called timelike, spacelike or null if $x^2$ is positive, negative or zero respectively. For timeline vectors $\sqrt{x^2}$ may be called the proper time $\tau$. The causal relation in $\mathbb{M}^d$ is $x \preccurlyeq y \Leftrightarrow y_0 \geq x_0$ and $(y - x)^2 \geq 0$.

The volume of the causal interval between $x \preccurlyeq y$ can be expressed in terms of the proper time between them. In $\mathbb{M}^d$ we have [2]

$$\mathrm{Vol}(y - x) = c_d((y - x)^2)^{d/2} \text{ with constant}$$
$$c_d = \frac{\pi^{(d-1)/2}}{2^{d-1} d\Gamma((d+1)/2)}, \quad (2)$$

where $\Gamma(z)$ is the Gamma function. Note, $c_2 = \frac{1}{2}$, $c_3 = \frac{\pi}{12}$, $c_4 = \frac{\pi}{24}$.

[*]Contact author: steven.p.johnston@gmail.com

We can generate a causal set by sprinkling into Minkowski spacetime. We place points according to a Poisson process such that the expected number of points in a region of volume $V$ is $\rho V$ where $\rho$ is a dimensionful parameter called the "sprinkling density." Having sprinkled the points we generate a causal set in which the elements are the sprinkled points and the causal relation is "read off" from the manifold's causal relation restricted to the sprinkled points.

## III. EMBEDDING REVIEW

To compare causal sets to Minkowski spacetime we use the notion of an embedding. An "embedding" of a causal set $(\mathcal{C}, \preccurlyeq)$ into a $\mathbb{M}^d$ is a map $p: \mathcal{C} \to \mathbb{M}^d$ which preserves the causal relations,

$$v_x \preccurlyeq v_y \text{ in } \mathcal{C} \Leftrightarrow p(v_x) \preccurlyeq p(v_y) \text{ in } \mathrm{M}. \tag{3}$$

A "faithful embedding" into $\mathbb{M}^d$ is an embedding such that the images of the causal set elements are uniformly distributed in $\mathbb{M}^d$. Further we require that the characteristic scale over which the manifold's geometry varies is much larger than the embedding scale.

For an embedding $p: \mathcal{C} \to \mathbb{M}^d$ we can think of this as two mappings $p_0: \mathcal{C} \to \mathbb{R}$ for the time coordinate and $\mathbf{p}: \mathcal{C} \to \mathbb{R}^{d-1}$ for the spatial coordinates, that is, $p(v_x) = (p_0(v_x), \mathbf{p}(v_x)) \in \mathbb{M}^d$.

We shall follow the framework in [5] with the causal set $(\mathcal{C}, \preccurlyeq)$ being assumed to be an $n$-element causal interval contained between a minimal element $v_1$ and a maximal element $v_n$: $\mathcal{C} = [v_1, v_n]$. If we seek to embed a causal set that is not an interval, we can restrict to intervals within it and embed them individually, combining and aligning the piecewise embeddings afterward.

We embed the minimal element $v_1$ at the origin of $\mathbb{M}^d$, $p(v_1) = (0, \mathbf{0})$, and the maximal element $v_n$ to a point on the time axis, $p(v_n) = (T, \mathbf{0})$ for some real value $T$ with dimensions of length $[T] = L$.

We pick a value of $T$ that is consistent with our causal set density $\rho$ such that $c_d T^d = \frac{n}{\rho}$ [compare (2) and (4)]. Picking a value for either $T$ or $\rho$ fixes the overall spatial scale of the causal set.

For causally related elements $v_x \prec v_y$ we can use the volume of their causal interval to define a causal set analog of their Minkowski proper time as

$$\tau(v_x, v_y) \coloneqq \left( \frac{|[v_x, v_y]|}{\rho c_d} \right)^{(1/d)}. \tag{4}$$

For convenience we also define $\tau(v_x, v_x) \coloneqq 0$ and $\tau(v_y, v_x) \coloneqq \tau(v_x, v_y)$. This proper time has the correct dimensions of length $[\tau(v_x, v_y)] = L$ and, by (2), we expect it to approximate the Minkowski norm,

$$\tau(v_x, v_y)^2 \approx (p(v_y) - p(v_x))^2. \tag{5}$$

We note that $\tau(v_1, v_n) = T$.

Note that some past work has suggested the length of the longest chain as a measure of the proper time between causally related elements (see [2] for an overview). This is appealing since the definition does not depend on $d$. However, we follow the volume-based approach here since it gives a better approximation with less statistical noise.

Following [5] we can use this $\tau$ to define a time coordinate for all $v_x \in \mathcal{C}$ as

$$t(v_x) \coloneqq \frac{T}{2} + \frac{\tau(v_1, v_x)^2 - \tau(v_x, v_n)^2}{2T} \tag{6}$$

such that

$$t(v_x) \approx p_0(v_x). \tag{7}$$

## IV. SPATIAL DISTANCES

We have so far recapped previous work from [5]. We now define a new function for the analog of the Euclidean spatial distance for two causally related elements $v_x \preccurlyeq v_y$ or $v_y \preccurlyeq v_x$ [compare (1)],

$$r(v_x, v_y) \coloneqq \sqrt{|(t(v_x) - t(v_y))^2 - \tau(v_x, v_y)^2|}. \tag{8}$$

We use the absolute value under the square root to ensure a non-negative value.

This spatial distance has the correct dimensions of length $[r(v_x, v_y)] = L$, is only defined for causally related pairs of elements, is symmetric $[r(v_x, v_y) = r(v_y, v_x)]$, and satisfies $r(v_x, v_x) = 0$. From (5) and (7) we expect that this approximates the Euclidean norm for causally related points,

$$r(v_x, v_y) \approx |\mathbf{p}(v_x) - \mathbf{p}(v_y)| \quad \text{and} \quad r(v_1, v_x) \approx |\mathbf{p}(v_x)|. \tag{9}$$

We would like to extend this spatial distance function to spacelike separated pairs of elements. Past work looking at spatial distances has often focused on the Minkowski norm distances between spacelike elements [2–4]. Our approach will be to use the time coordinate (6) to allow us to work just with the Euclidean norms rather than the Minkowski norms. In effect, we project from $\mathbb{M}^d$ to $\mathbb{R}^{d-1}$ by ignoring the time component of the vectors. This will simplify things considerably.

To make progress on the spatial distance for spacelike elements, we start by observing that the Euclidean norm satisfies the triangle inequality. For three Euclidean points $\mathbf{a}, \mathbf{b}, \mathbf{c} \in \mathbb{R}^n$ this is

$$|\mathbf{c} - \mathbf{a}| \leq |\mathbf{c} - \mathbf{b}| + |\mathbf{b} - \mathbf{a}|. \tag{10}$$

This inequality implies constraints between the spatial distances $r(v_x, v_y)$ which can be used to derive an approximate spatial distance for spacelike pairs of elements.

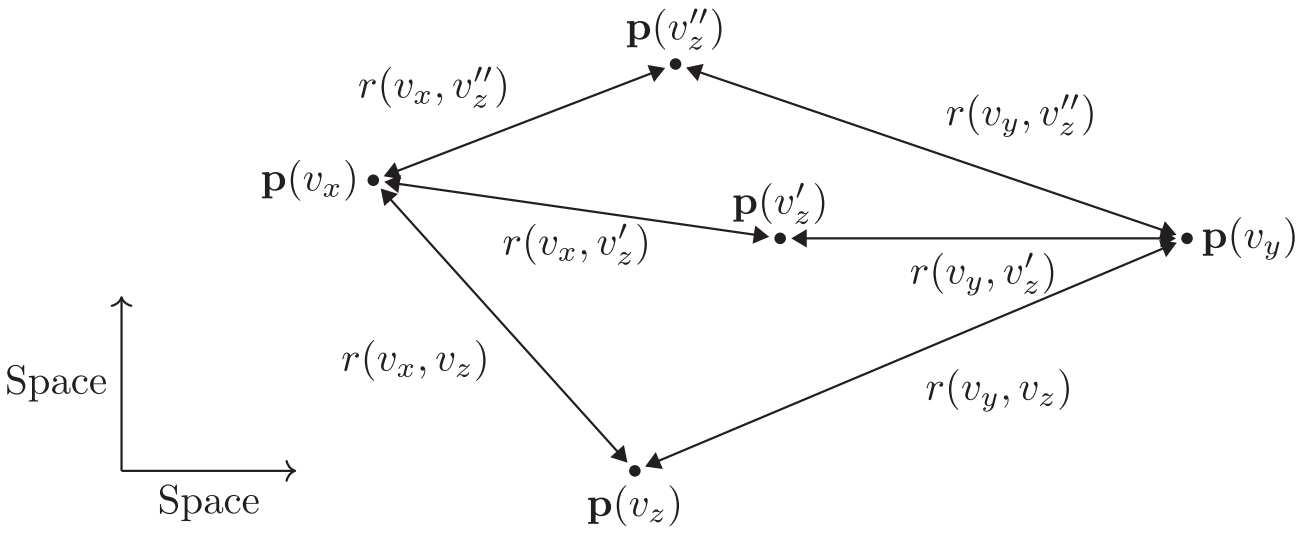

FIG. 1. An example two-dimensional spatial embedding (e.g., for embedding into $1+2$-dimensional Minkowski spacetime) showing just the *spatial* vectors for spacelike elements $v_x$ and $v_y$, along with three elements in their common past or future. Here the sum of $r(v_x, v_z') + r(v_y, v_z')$ gives a minimum value across $v_z, v_z', v_z''$.

For two spacelike elements $v_x || v_y$ we consider an element $v_z$ in their common past or future. Applying the triangle inequality we expect that

$$|\mathbf{p}(v_x) - \mathbf{p}(v_y)| \leq r(v_x, v_z) + r(v_y, v_z). \tag{11}$$

For different elements $v_z$ this bound will be tighter or looser. The tightest bound applies if the spatial components of the three embedded points are on a straight line $\mathbf{p}(v_x) \to \mathbf{p}(v_z) \to \mathbf{p}(v_y)$ (see Fig. 1). On a large causal set there will be many possible $v_z$, so it is likely there will be some $v_z$ that are close to this lower bound.

This suggests that if we minimize $r(v_x, v_z) + r(v_y, v_z)$ over all $v_z$ in the common past or future of $v_x$ and $v_y$ we will find a useful approximation for $|\mathbf{p}(v_x) - \mathbf{p}(v_y)|$.

For two spacelike elements $v_x || v_y$ their common past or future is the set of elements

$$S(v_x, v_y) = \{v_z \in \mathcal{C} | (v_x \preccurlyeq v_z \quad \text{and} \quad v_y \preccurlyeq v_z) \quad \text{or} \quad (v_z \preccurlyeq v_x \quad \text{and} \quad v_z \preccurlyeq v_y)\}. \tag{12}$$

Note that, since they are spacelike separated, and the causal relation is transitive, there are no elements in the past of one and in the future of the other. Also since $\mathcal{C}$ has a maximal and minimal element, $S(v_x, v_y)$ is always nonempty.

For $v_x || v_y$ we can then define (see Fig. 2)

$$r(v_x, v_y) \coloneqq \min_{v_z \in S(v_x, v_y)} r(v_x, v_z) + r(v_y, v_z). \tag{13}$$

This is well defined since each $r(v_x, v_z)$ and $r(v_y, v_z)$ are defined for causally related pairs and the minimization is over a finite number of elements.

We can simplify the minimization if we define a matrix

$$R_{xy} = \begin{cases} r(v_x, v_y) & \text{if } v_x \preccurlyeq v_y \text{ or } v_y \preccurlyeq v_x, \\ \infty & \text{if } v_x || v_y. \end{cases} \tag{14}$$

Then for $v_x || v_y$ we can minimize the sum over all $v_z \in \mathcal{C}$,

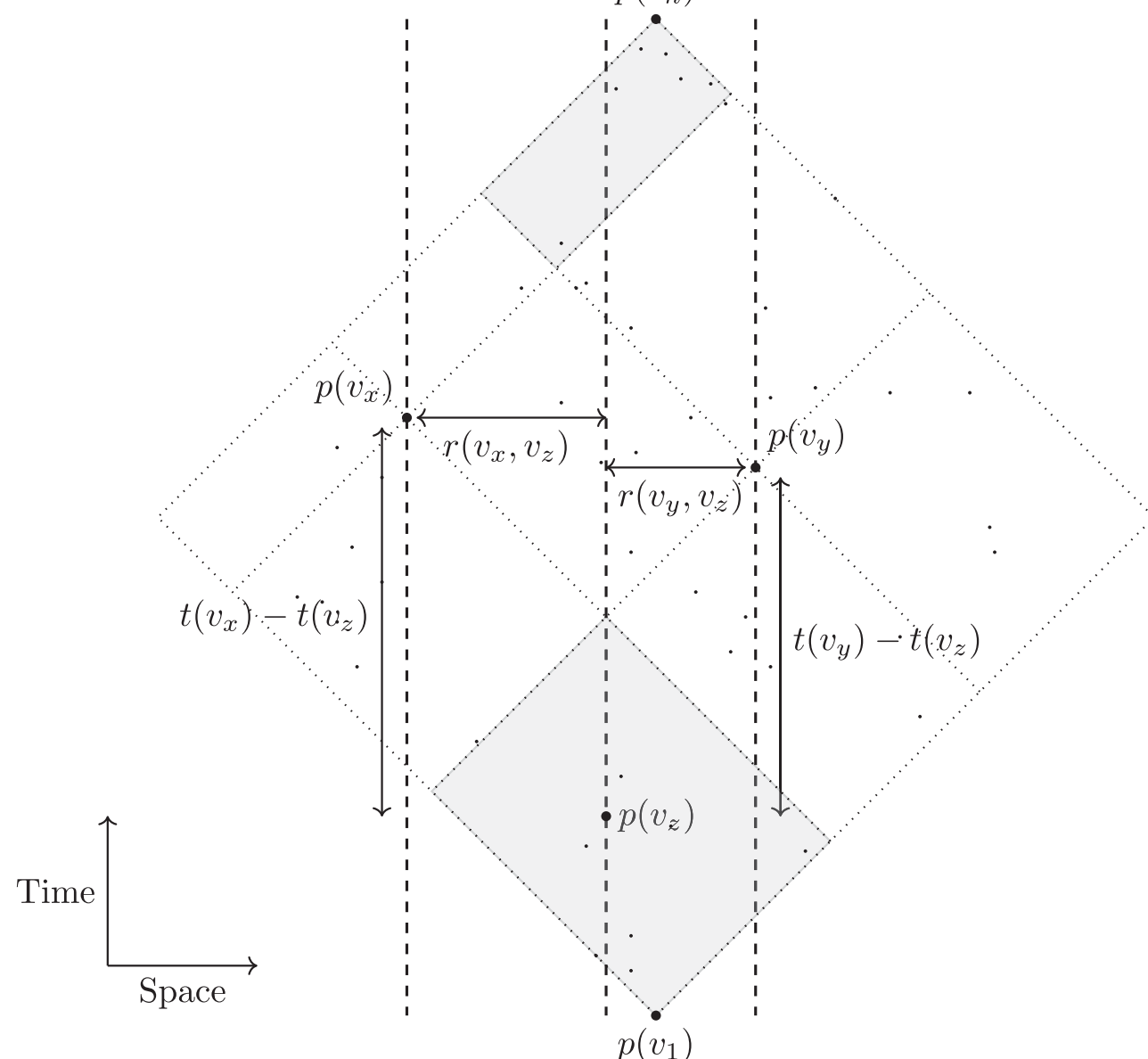

FIG. 2. A $1+1$-dimensional spacetime diagram for the embeddings of spacelike elements $v_x$ and $v_y$, along with $v_z$ in their common past. The spatial distances $r(v_x, v_z)$ and $r(v_y, v_z)$ are shown and the set $S(v_x, v_y)$ is shaded. Since this only shows one spatial dimension, the sum of $r(v_x, v_z) + r(v_y, v_z)$ appears equal to $r(v_x, v_y)$. In higher dimensions this equality or inequality depends on the spatial arrangement of the $\mathbf{p}(v_x), \mathbf{p}(v_y), \mathbf{p}(v_z)$ vectors (see Fig. 1).

$$r(v_x, v_y) = \min_{v_z \in \mathcal{C}} (R_{xz} + R_{yz}). \tag{15}$$

This minimization is equivalent to calculating the matrix $R^2$ using min-plus algebra for the matrix multiplication.

It is interesting to note that minimizing the sum of piecewise spatial distances between $v_x$ and $v_y$ is similar to the shortest-distance property of straight lines or, more generally, geodesics. This offers some encouragement that this minimization approach may generalize for embeddings into curved spacetimes.

With (8) for causally related pairs and (15) for spacelike pairs, we have now defined the $r(v_x, r_y)$ function for all pairs of elements, with the expectation that $r(v_x, v_y) \approx |\mathbf{p}(v_x) - \mathbf{p}(v_y)|$.

## V. EMBEDDINGS

Now that we have Euclidean spatial distances for *all* pairs of elements, we are able to apply standard embedding methods such as multidimensional scaling (MDS). This uses a decomposition of the distance matrix to find $(d-1)$-dimensional spatial vectors which recreate these distances. These vectors can then serve as the embedding $\mathbf{p}(v_x)$.

In [7] they applied multidimensional scaling using Minkowski norm distances, following the method in [2] to approximate Minkowski norms for spacelike pairs of elements. The use of Minkowski norms, rather than Euclidean norms in that approach required modifications to the standard MDS approach to deal with the indefinite metric signature. Since our distances $r(v_x, v_y)$ are Euclidean spatial distances, the standard MDS approach can be applied.

To follow the MDS approach (for an overview, see [7]), we define a $(d-1) \times n$ matrix $P$ where the $x$th column will be the vector $\mathbf{p}(v_x)$.

Each entry in the $n \times n$ matrix $P^T P$ is an inner product: $(P^T P)_{xy} = \mathbf{p}(v_x) \cdot \mathbf{p}(v_y)$. We can express this in terms of Euclidean norms via a polarization identity,

$$\mathbf{p}(v_x) \cdot \mathbf{p}(v_y) = \frac{1}{2}(|\mathbf{p}(v_x)|^2 + |\mathbf{p}(v_y)|^2 - |\mathbf{p}(v_x) - \mathbf{p}(v_y)|^2). \tag{16}$$

The equivalent expression for the causal set is the matrix

$$X_{xy} \coloneqq \frac{1}{2}(r(v_1, v_x)^2 + r(v_1, v_y)^2 - r(v_x, v_y)^2) \tag{17}$$

with the expectation that $X_{xy} \approx \mathbf{p}(v_x) \cdot \mathbf{p}(v_y)$.

The $n \times n$ symmetric matrix $X$ is defined for all $v_x, v_y \in \mathcal{C}$, whether causally related or spacelike separated. By contrast in [5] we constructed a rectangular matrix of inner products only defined for causally related pairs of elements. This led to more complexity and a multistage embedding procedure, first for a light cone of mutually related elements, then for the remaining spacelike elements. Here we shall obtain the embedding for all the causal set elements at once.

We can derive the matrix $P$ from $X$ by using its eigendecomposition. We write

$$X = U^T \Sigma U \tag{18}$$

for a matrix of eigenvectors $U$ and a diagonal matrix of eigenvalues $\Sigma$. Since $X$ is symmetric and real, its eigenvalues are real.

Typically we expect $r(v_x, v_y)$ to be such that $\mathrm{rank}(X) \geq d-1$. Since we seek $(d-1)$-dimensional vectors we can use a truncated eigendecomposition where we keep the $d-1$ largest positive eigenvalues,

$$X \approx U_{d-1}^T \Sigma_{d-1} U_{d-1}. \tag{19}$$

We then define

$$P \coloneqq \sqrt{\Sigma_{d-1}} U_{d-1}, \tag{20}$$

where we have kept the $d-1$ largest positive eigenvalues and their eigenvectors. Then we expect $P^T P \approx X \approx \mathbf{p}(v_x) \cdot \mathbf{p}(v_y)$. The square root introduces some sign freedom for the matrix $P$. Picking $\pm$ signs for each diagonal entry in $\sqrt{\Sigma_{d-1}}$ is equivalent to a choice for spatial parity for each spatial dimension. This is similar to the sign freedom for parity in [5].

This gives us the embedding vectors $\mathbf{p}(v_x)$ as columns of the $P$ matrix. Combining this with the time coordinates (6) we have found our full embedding $p(v_x) = (t(v_x), \mathbf{p}(v_x)) \in \mathbb{M}^d$.

## VI. SIMULATIONS

One way to assess the accuracy of our embedding $p : \mathcal{C} \to \mathbb{M}^d$ is to perform simulations where we generate a causal set by sprinkling, calculate the embedding and compare the results to the original sprinkled points.

Our simulation procedure is as follows:

(1) Fix a dimension $d$ and sprinkling density $\rho$.

(2) Define a unit interval $\mathbb{I}$ in $\mathbb{M}^d$ from (0,0) to (1,0). This has volume $c_d$. Pick the total number of points to sprinkle based on a Poisson distribution with mean $c_d \rho$.

(3) Sprinkle these points into $\mathbb{I}$ by choosing random coordinates within $\mathbb{I}$. Add in $(0, \mathbf{0})$ and $(1, \mathbf{0})$. Call the resulting points $P_x$ for $x = 1, \ldots, n$.

(4) Generate a causal set from the sprinkled points $P_x$ by reading off their causal relations. Call this $(\mathcal{C}, \preccurlyeq)$.

(5) Calculate the embedding $p : \mathcal{C} \to \mathbb{M}^d$ using the methods presented above. Note that the embedding only depends on relations in the causal set, not on the original sprinkled points.

There is no guarantee that the original sprinkled points $P_x$ and the embedded points $p(v_x)$ will share the same spatial frame of reference. In general there will be an orthogonal transformation that is needed to align these frames of reference. This can be calculated by solving the orthogonal procrustean problem for the two sets of points (see [5] for details). The resulting orthogonal matrix can be applied to the spatial vectors $\mathbf{p}(v_x)$ to allow a direct comparison of the coordinates $P_x$ and $p(v_x)$.

## VII. RESULTS

After a simulation we have two sets of aligned points $P_x$ and $p(v_x)$. If the embedding perfectly recovered the sprinkling, we would expect $P_x = p(v_x)$. In practice we find that $P_x \approx p(v_x)$ and we want to assess the quality of this approximation. To this end we compare (see [5] for more details):

(i) the causal relations between $\mathcal{C}$ and the $p(v_x)$ points, see Table I;

(ii) the causal interval volumes $I(v_x, v_y)$ and $\mathrm{Vol}(P_y - P_x)$ for all $v_x \preccurlyeq v_y$, see Table II;

(iii) the original coordinates $P_x$ and the embedded coordinates $p(v_x)$, see Table III;

(iv) the Minkowski norm distances $(P_x - P_y)^2$ and $(p(v_x) - p(v_y))^2$ for all pairs of $v_x, v_y \in \mathcal{C}$, including distances for spacelike separated points, see Table IV; and

(v) the spatial distances $|\mathbf{P}_x - \mathbf{P}_y|$ and $|\mathbf{p}(v_x) - \mathbf{p}(v_y)|$ for all pairs of $v_x, v_y \in \mathcal{C}$, including distances for spacelike separated points, see Table V.

We show some example embeddings in Figs. 3–8.

## VIII. CONCLUSIONS AND FURTHER WORK

The simulation results show that the derived embeddings are high quality, with correlations between the sprinkled values and the embedded values consistently reaching

TABLE I. Mean and standard deviation (std.) of sensitivity and specificity for ten simulations for $d = 2, 3, 4$ and $n = 500, 1000, 1500$.

| Causal relations | | | |
|---|---|---|---|
| $d = 2$ | 500 | 1000 | 1500 |
| Mean sensitivity | 0.9824 | 0.9832 | 0.9856 |
| Std. sensitivity | 0.0023 | 0.0025 | 0.0020 |
| Mean specificity | 0.9823 | 0.9837 | 0.9859 |
| Std. specificity | 0.0033 | 0.0029 | 0.0025 |
| $d = 3$ | 500 | 1000 | 1500 |
| Mean sensitivity | 0.9767 | 0.9786 | 0.9807 |
| Std. sensitivity | 0.0051 | 0.0042 | 0.0058 |
| Mean specificity | 0.9847 | 0.9895 | 0.9896 |
| Std. specificity | 0.0039 | 0.0027 | 0.0027 |
| $d = 4$ | 500 | 1000 | 1500 |
| Mean sensitivity | 0.9210 | 0.9467 | 0.9642 |
| Std. sensitivity | 0.0321 | 0.0228 | 0.0101 |
| Mean specificity | 0.9790 | 0.9869 | 0.9897 |
| Std. specificity | 0.0056 | 0.0023 | 0.0025 |

TABLE II. Mean and standard deviation of correlation for causal interval volumes for ten simulations for $d = 2, 3, 4$ and $n = 500, 1000, 1500$.

| Volume correlation | | | |
|---|---|---|---|
| $d = 2$ | 500 | 1000 | 1500 |
| Mean correlation | 0.9968 | 0.9976 | 0.9982 |
| Std. correlation | 0.0010 | 0.0005 | 0.0002 |
| $d = 3$ | 500 | 1000 | 1500 |
| Mean correlation | 0.9943 | 0.9966 | 0.9973 |
| Std. correlation | 0.0010 | 0.0005 | 0.0007 |
| $d = 4$ | 500 | 1000 | 1500 |
| Mean correlation | 0.9815 | 0.9904 | 0.9943 |
| Std. correlation | 0.0033 | 0.0022 | 0.0011 |

TABLE III. Mean and standard deviation of correlation of $P_x$ and $p(v_x)$ coordinates for ten simulations for $d = 2, 3, 4$ and $n = 500, 1000, 1500$.

| Coordinate correlation | | | | |
|---|---|---|---|---|
| $d = 2$, $n = 500$ | $x_0$ | $x_1$ | | |
| Mean correlation | 0.999 | 0.9948 | | |
| Std. correlation | 0.0006 | 0.0027 | | |
| $d = 2$, $n = 1000$ | $x_0$ | $x_1$ | | |
| Mean correlation | 0.9996 | 0.9961 | | |
| Std. correlation | 0.0001 | 0.0012 | | |
| $d = 2$, $n = 1500$ | $x_0$ | $x_1$ | | |
| Mean correlation | 0.9997 | 0.9975 | | |
| Std. correlation | 0.0001 | 0.0004 | | |
| $d = 3$, $n = 500$ | $x_0$ | $x_1$ | $x_2$ | |
| Mean correlation | 0.9979 | 0.987 | 0.9887 | |
| Std. correlation | 0.0009 | 0.0038 | 0.0014 | |
| $d = 3$, $n = 1000$ | $x_0$ | $x_1$ | $x_2$ | |
| Mean correlation | 0.999 | 0.9949 | 0.9952 | |
| Std. correlation | 0.0002 | 0.0018 | 0.0011 | |
| $d = 3$, $n = 1500$ | $x_0$ | $x_1$ | $x_2$ | |
| Mean correlation | 0.9993 | 0.9966 | 0.9966 | |
| Std. correlation | 0.0003 | 0.0014 | 0.0015 | |
| $d = 4$, $n = 500$ | $x_0$ | $x_1$ | $x_2$ | $x_3$ |
| Mean correlation | 0.9941 | 0.9142 | 0.9319 | 0.9289 |
| Std. correlation | 0.0011 | 0.0414 | 0.0162 | 0.0131 |
| $d = 4$, $n = 1000$ | $x_0$ | $x_1$ | $x_2$ | $x_3$ |
| Mean correlation | 0.9972 | 0.9573 | 0.9661 | 0.9650 |
| Std. correlation | 0.0004 | 0.0123 | 0.0081 | 0.0082 |
| $d = 4$, $n = 1500$ | $x_0$ | $x_1$ | $x_2$ | $x_3$ |
| Mean correlation | 0.9982 | 0.9751 | 0.9778 | 0.9744 |
| Std. correlation | 0.0002 | 0.0052 | 0.0055 | 0.0057 |

TABLE IV. Mean and standard deviation of correlation for Minkowski distances, including spacelike separated points, for ten simulations for $d = 2, 3, 4$ and $n = 500, 1000, 1500$.

| Minkowski distance correlation | | | |
|---|---|---|---|
| $d = 2$ | 500 | 1000 | 1500 |
| Mean correlation | 0.9964 | 0.9978 | 0.9983 |
| Std. correlation | 0.001 | 0.0008 | 0.0003 |
| $d = 3$ | 500 | 1000 | 1500 |
| Mean correlation | 0.9912 | 0.9954 | 0.9970 |
| Std. correlation | 0.0014 | 0.0008 | 0.0006 |
| $d = 4$ | 500 | 1000 | 1500 |
| Mean correlation | 0.9578 | 0.9771 | 0.9851 |
| Std. correlation | 0.0183 | 0.0031 | 0.0018 |

TABLE V. Mean and standard deviation of correlation for spatial distances, including causally related and spacelike separated points, for ten simulations for $d = 2, 3, 4$ and $n = 500, 1000, 1500$.

| Spatial distance correlation | | | |
|---|---|---|---|
| $d = 2$ | 500 | 1000 | 1500 |
| Mean correlation | 0.9891 | 0.9931 | 0.9943 |
| Std. correlation | 0.0029 | 0.0015 | 0.0019 |
| $d = 3$ | 500 | 1000 | 1500 |
| Mean correlation | 0.9738 | 0.9855 | 0.9920 |
| Std. correlation | 0.0133 | 0.0044 | 0.0037 |
| $d = 4$ | 500 | 1000 | 1500 |
| Mean correlation | 0.7892 | 0.8874 | 0.9282 |
| Std. correlation | 0.0312 | 0.0129 | 0.0083 |

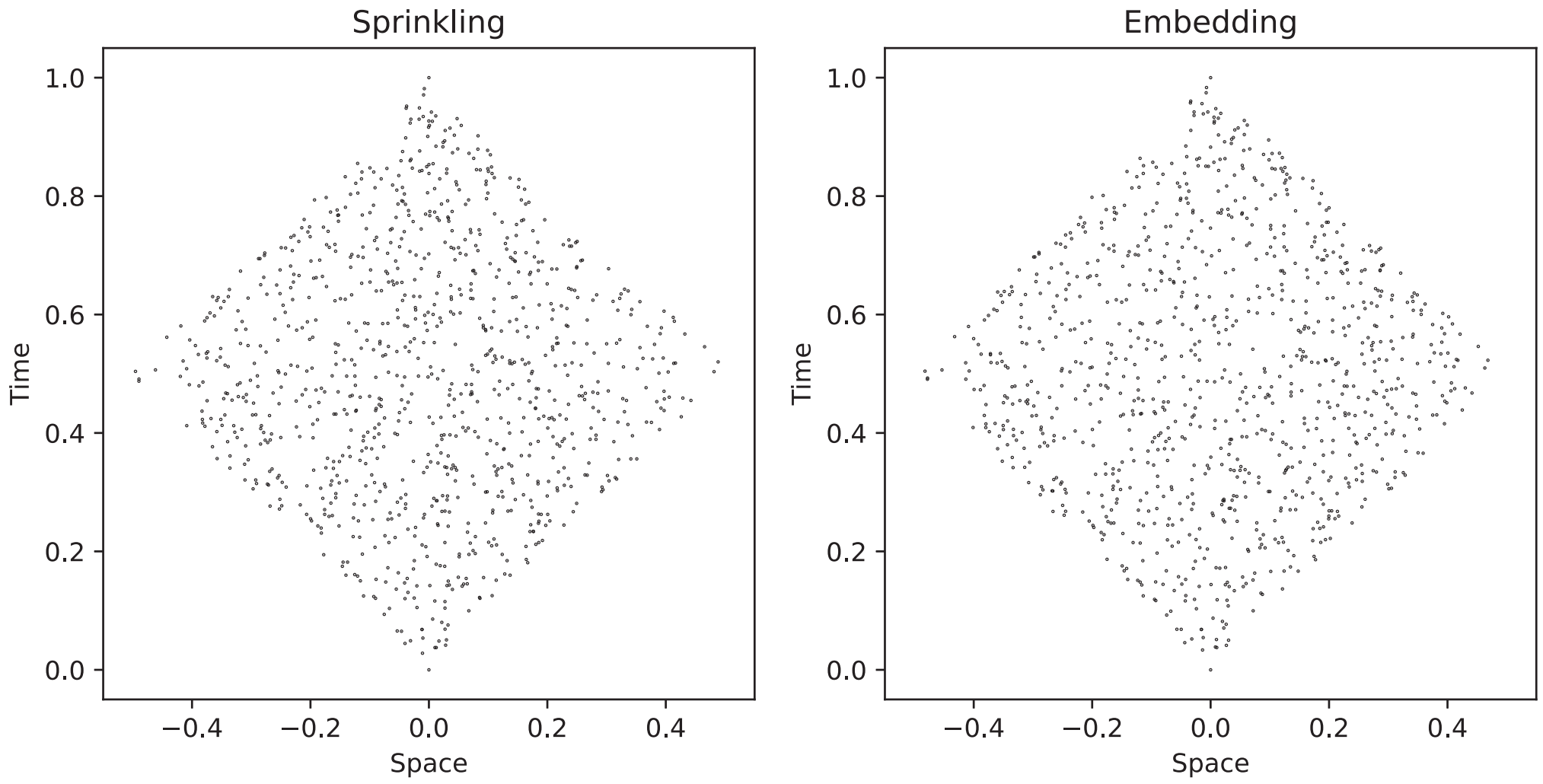

FIG. 3. Comparison of $n = 1000$, $d = 2$ sprinkling and the corresponding embedding.

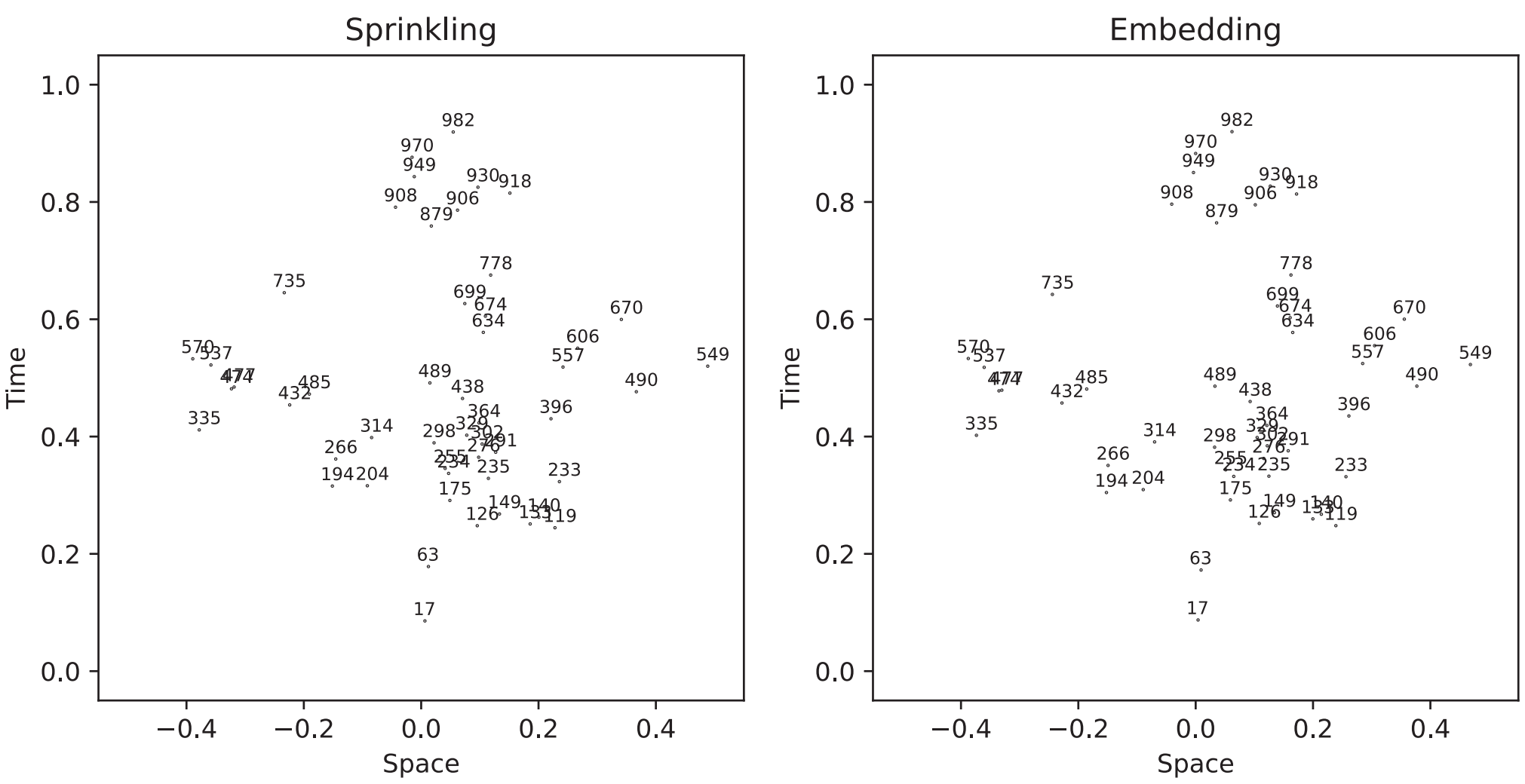

FIG. 4. Comparison of the same embedding with a random sample of the points shown and labeled for readability.

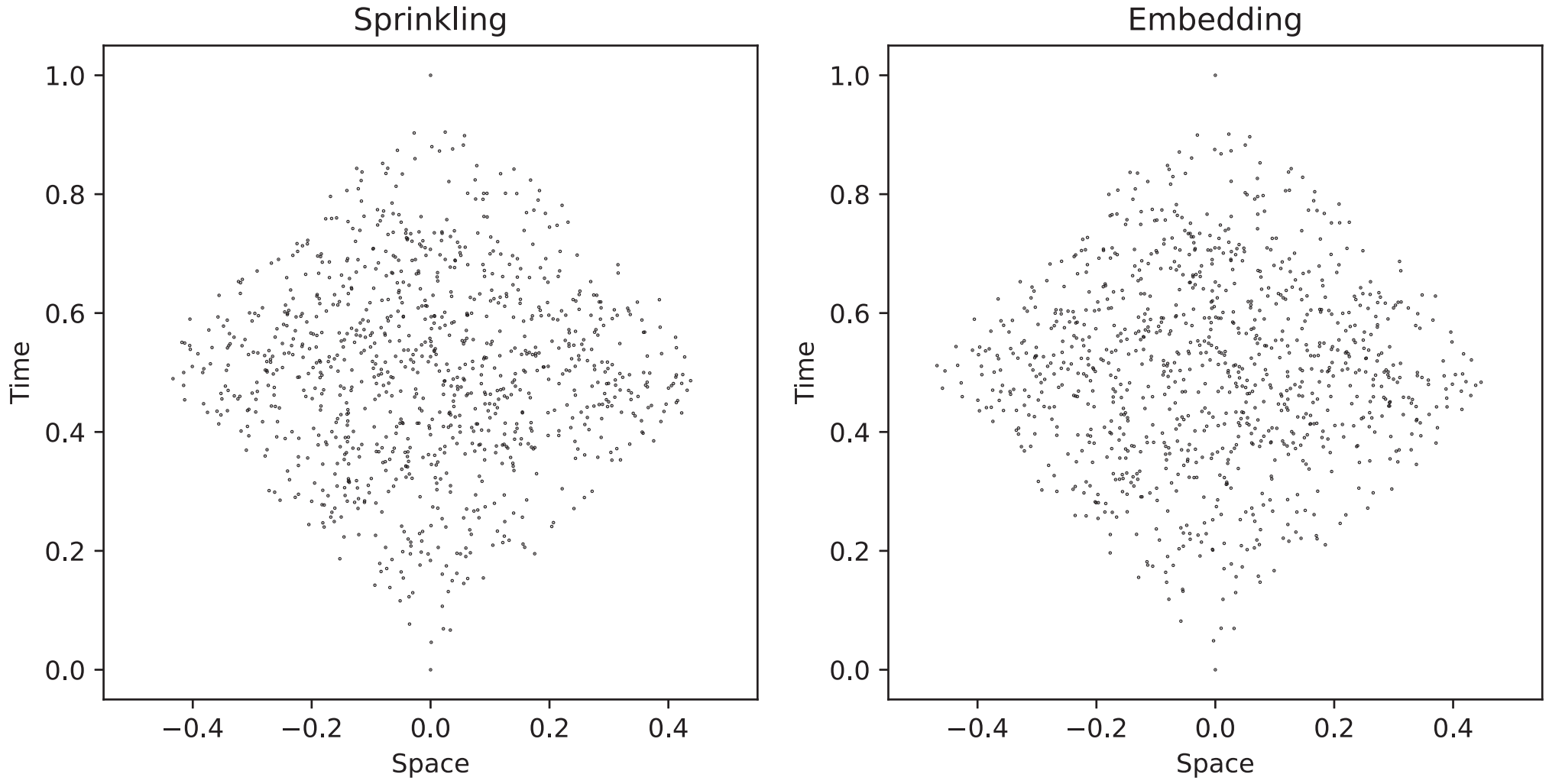

FIG. 5. Comparison showing the $x_0$ and $x_1$ coordinates for $n = 1000$, $d = 3$ sprinkling and the corresponding embedding.

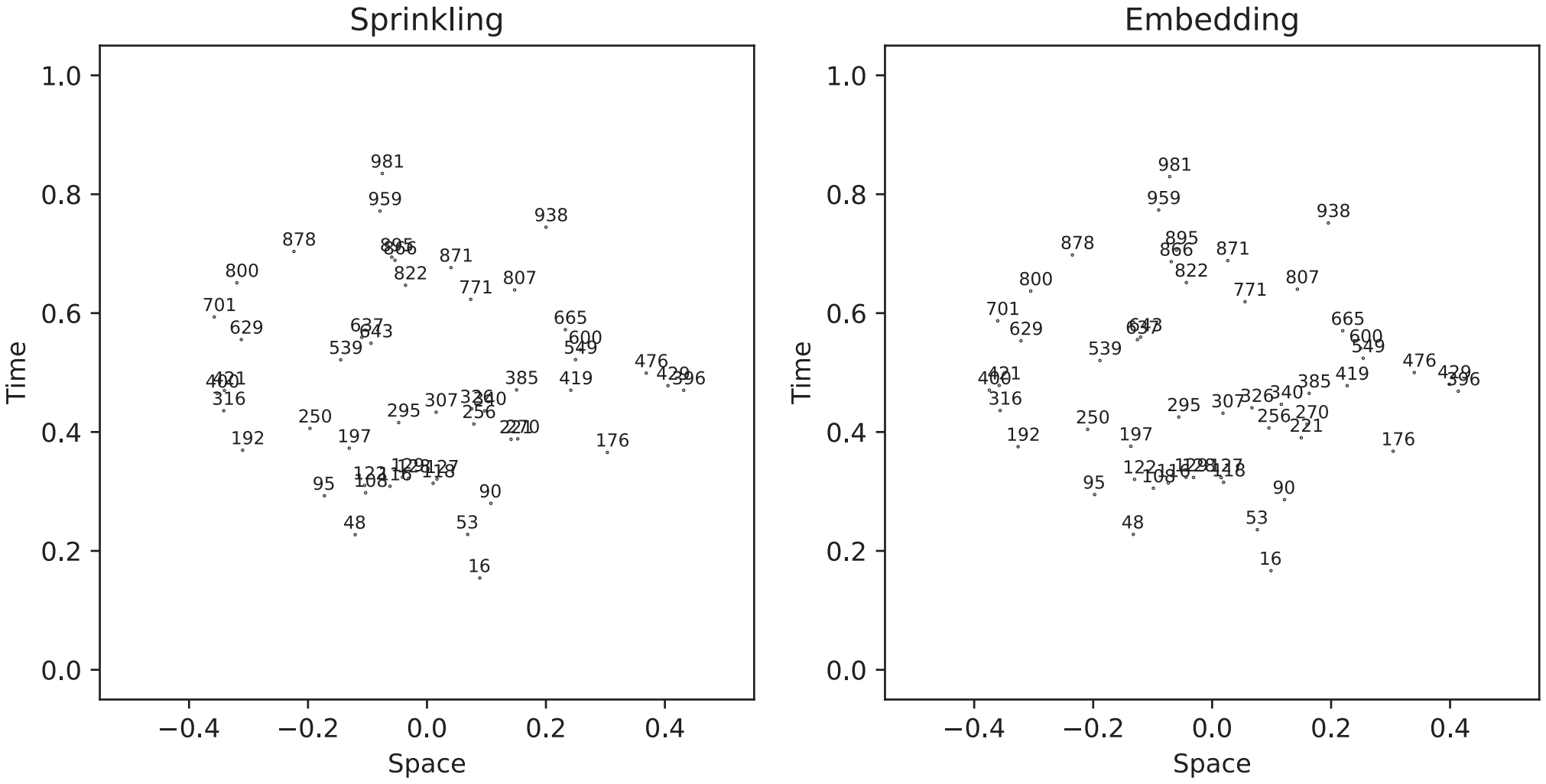

FIG. 6. Comparison of the same embedding with a random sample of the points shown and labeled for readability.

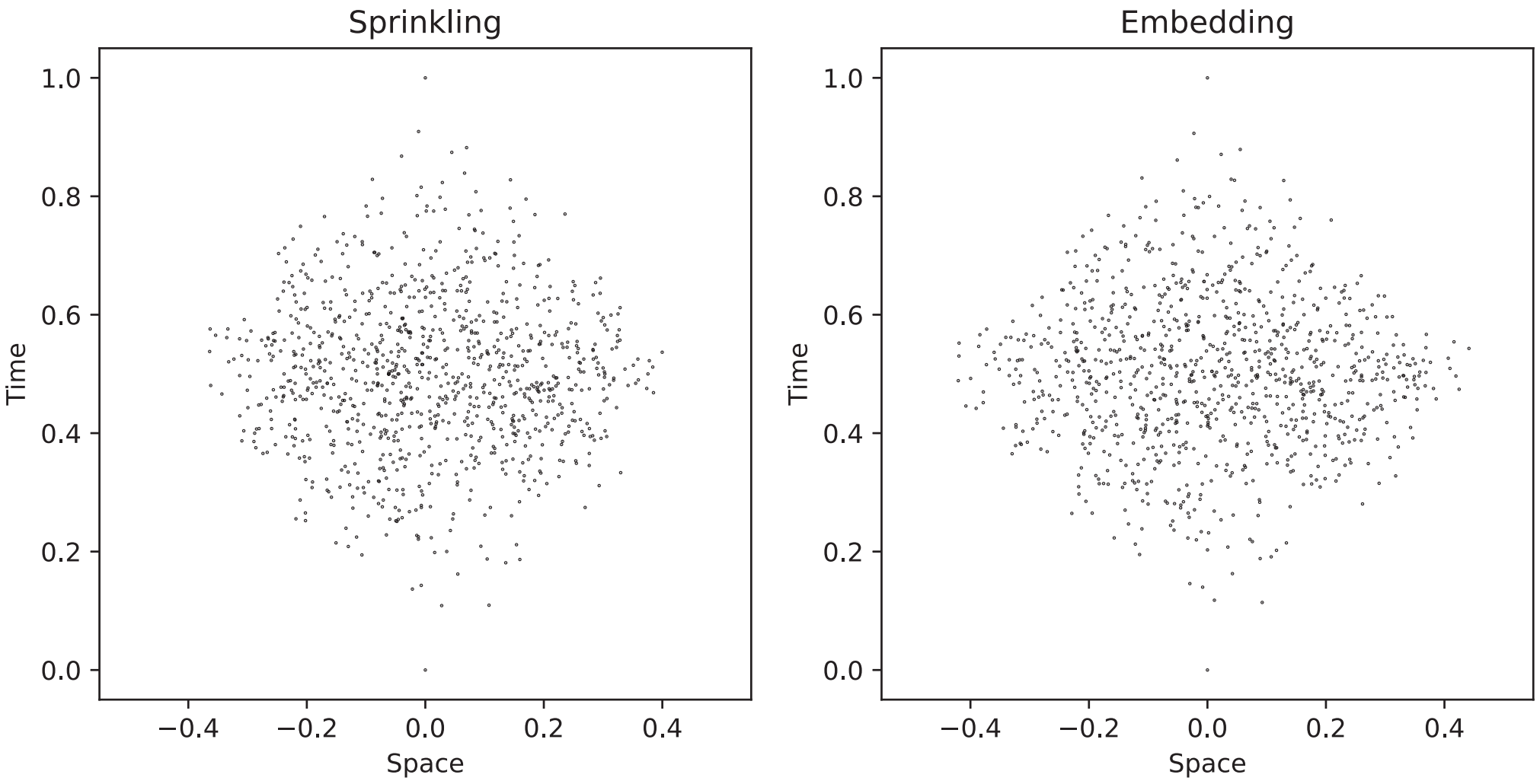

FIG. 7. Comparison showing the $x_0$ and $x_1$ coordinates for $n = 1000$, $d = 4$ sprinkling and the corresponding embedding.

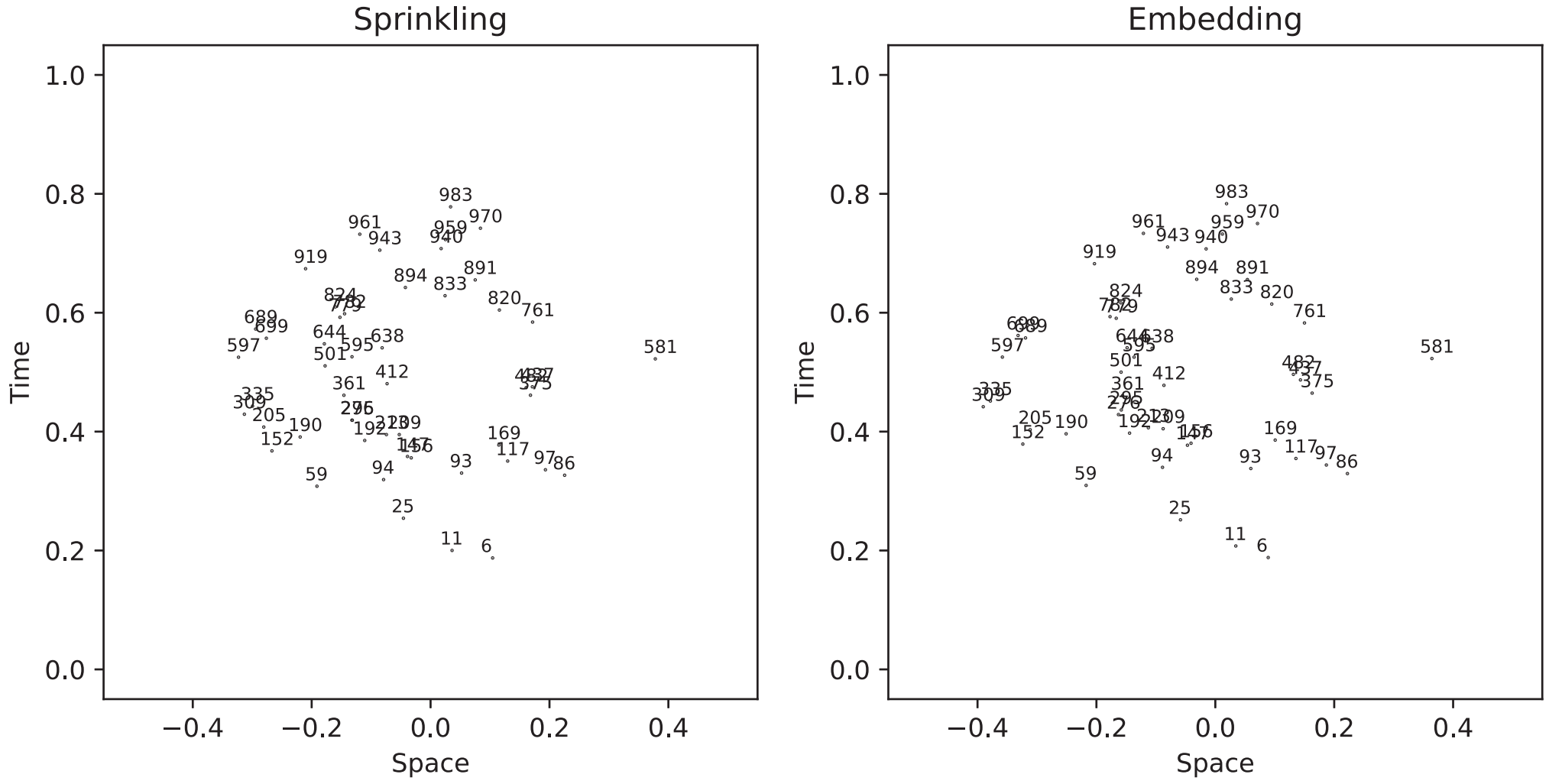

FIG. 8. Comparison of the same embedding with a random sample of the points shown and labeled for readability.

over 0.95. These are higher than the results in [5], with particular improvements for $d = 3$, 4 spatial coordinates and Minkowski norms. These results in $d > 2$ are encouraging given the relatively small size of the causal sets being simulated.

The approach is conceptually simple, using direct analogs of continuum geometric constructions (spacetime volumes, proper times) together with the familiar triangle inequality, to construct a matrix of spatial distance analogs. Since these are Euclidean distances, not Minkowski norms, the widely used apparatus of vector embeddings, such as MDS, can be directly applied.

The method has focused on embedding causal sets into Minkowski spacetimes. We note, however, that the method applies to any causal set, including those generated by sprinkling into curved spacetime. Further work to explore the quality of embeddings into curved spacetimes would be interesting. We hope that this method will open up new and simpler ways to explore analogs for continuum geometric constructions, helping to bridge the gap between discrete causal set theory and familiar theories using continuum spacetime backgrounds.

## DATA AVAILABILITY

The data supporting this study's findings are available within the article.